\documentclass[12pt]{article}

\usepackage{graphicx}

\newcommand{\tr}{\mbox{Tr} \, }
\newcommand{\ket}[1]{\left | #1 \right \rangle}
\newcommand{\bra}[1]{\left \langle #1 \right |}

\newcommand{\proj}[1]{\ket{#1} \! \bra{#1}}

\newcommand{\superop}{{\cal E}}
\newcommand{\errcorr}{{\cal D}}

\newcommand{\bigsum}[1]{{\displaystyle \sum_{#1}}}
\newcommand{\entform}{\mbox{\bf E}}

\begin{document}

\title{Entanglement and perfect quantum error correction}

\author{Benjamin Schumacher$^{(1)}$
        and Michael D. Westmoreland$^{(2)}$}
\maketitle
\begin{center}
{\sl
$^{(1)}$Department of Physics, Kenyon College, Gambier, OH 43022 USA \\
$^{(2)}$Department of Mathematical Sciences, Denison University,
 Granville, OH  43023 USA }
\end{center}

\section*{Abstract}

The entanglement of formation gives a necessary and sufficient condition
for the existence of a perfect quantum error correction procedure.

\section{Quantum error correction}

Suppose a composite quantum system $RQ$ is initially in a pure joint input
state $\ket{\Psi^{RQ}}$.  The subsystem $Q$ undergoes a dynamical evolution
described by a trace-preserving, completely positive (CP) map $\superop$.
The joint output state is therefore
\begin{equation}
	\rho^{RQ'} = I \otimes \superop \left ( \proj{\Psi^{RQ}} \right ).
\end{equation}
This situation describes the transmission of ``quantum information'' (the
entanglement between $R$ and $Q$) via a noisy quantum channel.  For example,
imagine that $RQ$ is a quantum computing device.
The overall state of the device is entangled.  Subsystem $Q$ is
imperfectly isolated from the environment, and thus experiences noise and
distortion given by $\superop$.  The problem of sending entanglement
through a channel in this way is closely related to other tasks of quantum
information transfer, such as the transmission of an unknown quantum state
of $Q$ \cite{s96}.

We are interested in the question of whether the original input
state $\ket{\Psi^{RQ}}$ can be restored by some possible operation on $Q$
alone.  Such a restoring operation is called a ``quantum error correction''
procedure \cite{mikeandike}.
We say that {\em perfect} quantum error correction is possible
when there exists a trace preserving CP map $\errcorr$ on $Q$ such that
\begin{equation}
	\proj{\Psi^{RQ}} = I \otimes \errcorr \left ( \rho^{RQ'} \right ) .
\end{equation}
If no such $\errcorr$ exists, we may still be able to do {\em approximate}
quantum error correction, in which case we restore $\rho^{RQ'}$ to a state
close to the original.  (``Close'' here is usually defined in terms of the
fidelity or some equivalent measure.)  In this paper, we will mostly be
concerned with the question of perfect (unit fidelity) error correction.

We first note that, if the input state of $RQ$ is a product state, then
it is always possible to restore the input state by means of an operation
on $Q$.  Thus, the problem of error correction is only non-trivial when
$\ket{\Psi^{RQ}}$ is entangled.  The entanglement of a pure state of
$RQ$ is measured by the entropy $S^Q$ of the subsystem $Q$:
\begin{equation}
	S^Q = - \tr \rho^Q \log \rho^Q .
\end{equation}
Of course, since $RQ$ is in a pure state, then $S^Q = S^R$.

The output state $\rho^{RQ'}$ is generally not pure.  Schumacher and
Nielsen \cite{sn96} defined the ``coherent information'' to be
\begin{equation}
	I = S^{Q'} - S^{RQ'} .
\end{equation}
This quantity has a number of significant properties.  It is positive only
if the output state $\rho^{RQ'}$ is entangled.  Furthermore, it cannot be
increased by any operation on $Q$ alone.  Since the initial coherent
information is just $S^Q$, this means that $I \leq S^Q$ after the action
of $\superop$.  Furthermore, any loss of $I$ due to the action of $\superop$
is irreversible, i.e., cannot be reversed by any subsequent evolution of
$Q$.  It follows that $I = S^Q$ is a necessary condition for the existence
of a perfect quantum error correction operation $\errcorr$.

In \cite{sn96},
it is shown that the condition $I = S^Q$ is also {\em sufficient} for
the existence of such an operation.  In outline, we imagine a larger quantum
system $RQE$ that includes the environment $E$ with which $Q$ interacts.
The initial state of the environment is a pure state $\ket{0^E}$, and
the interaction of $Q$ and $E$ is described by the unitary operator
$U^{QE}$.  (Since the operation $\superop$ is a trace-preserving CP map,
it must always be realizable in this way as a unitary evolution on a
larger system.)  The condition $I = S^Q$ implies that the output state of
the subsystem $RE$ is a product state.  From this product structure,
a perfect error correction procedure can be constructed.  In short,
the lack of any correlation between $R$ and $E$ after the evolution is
sufficient to permit the restoration of the original $QE$ state by an
error-correction operation $\errcorr$.

\section{Entanglement of formation}

The coherent information $I$ is a measure of the entanglement of $Q$ with $R$
after it has undergone its noisy evolution.  There are, however, many other
ways to measure the entanglement of the output state $\rho^{RQ'}$.
One of the most fundamental is the
``entanglement of formation'' \cite{entangle},
denoted $\entform$.  The entanglement of
formation of a pure state $\ket{\psi^{AB}}$ is just $\entform = S^A$, the
entropy of one of the subsystems.  A mixed state $\rho^{AB}$ has an
entanglement of formation
\begin{equation}
	\entform = \min \sum_{k} p_k \entform_k
\end{equation}
where $\entform_k$ is the entanglement of the pure state $\ket{\phi^{AB}_k}$ and
the minimum is taken over all pure-state ensembles such that $\rho^{AB} =
\bigsum{k} p_k \proj{\phi^{AB}_k}$.  $\entform$ has the property
that it cannot be increased by local quantum operations on, or the exchange
of classical information between, the two subsystems.

The entanglement of formation is related to the ``entanglement resources''
necessary to create the quantum state.  However, to make this connection
sharp one must define an {\em asymptotic} entanglement of formation
\begin{equation}
	\entform_{\infty} ( \rho^{AB} ) = \lim_{n \rightarrow \infty}
		\frac{1}{n} \entform \left (  (\rho^{AB})^{\otimes n} \right ).
\end{equation}
$\entform_{\infty}$ is the asymptotic number of maximally entangled qubit
pairs needed to create the state $\rho^{AB}$ by local operations and classical
communication---that is, for large $n$, about $n \entform_{\infty}$ pairs
are required to make $n$ copies of the state $\rho^{AB}$.  Though the definitions
of $\entform_{\infty}$ and $\entform$ are distinct, and we can see that
$\entform_{\infty} \leq \entform$, it is not known whether or not these
are actually equal in general \cite{wkw1998}.
We will here use the ``single system''
definition of the entanglement of formation $\entform$, since we are not
primarily concerned with asymptotic questions.

The coherent information $I$ and the entanglement of formation $\entform$
of the output state $\rho^{RQ'}$ satisfy $I \leq \entform$.
To see this, suppose we have an
ensemble of $RQ$ states such that $\rho^{RQ'} = \bigsum{k} p_k \rho^{RQ}_k$,
then
\begin{equation}
	S^{Q'} - \sum_k p_k S^{Q}_k \leq S^{RQ'} - \sum_k p_k S^{RQ}_k .
\end{equation}
(This follows from the strong subadditivity of the entropy functional
\cite{sww}.)
For any ensemble of pure states, $S^{RQ}_k = 0$ and so
\begin{eqnarray}
	S^{Q'} - S^{RQ'} \leq \sum_k p_k S^Q_k .
\end{eqnarray}
If we choose the pure state ensemble that minimizes the right-hand side,
we obtain $I \leq \entform$.

For the input pure state of $RQ$, both $\entform$ and $I$ are equal to $S^Q$.
For the output state, the condition that $\entform = S^Q$ is weaker than
the condition that $I = S^Q$, since we can have $I < \entform$.
Thus, $\entform = S^Q$ is a necessary condition for the existence
of a perfect quantum error correction operation $\errcorr$.  Remarkably, it
turns out that this is also a sufficient condition.  We now show this.
Suppose that $\entform = S^Q$ for our output state $\rho^{RQ'}$.
Our argument is based on three facts.

{\bf Fact 1:  Concavity of the entropy.}
Suppose we write a mixed state as an ensemble of states:
$\rho = \bigsum{k} p_k \rho_k$.  Then
\begin{equation}
	S \geq \sum_k p_k S_k   \label{concavity}
\end{equation}
with equality if and only if $\rho_k = \rho$ for all $k$ with $p_k > 0$
\cite{wehrl}.
In our context, the condition that $\entform = S^Q$ means that
\begin{equation}
	0 = S^{R'} - \min \sum_k p_k S^R_k ,
\end{equation}
where the minimum is taken over all pure state ensembles for $\rho^{RQ'}$.
Equation~\ref{concavity} then tells us that
\begin{equation}
	0 = S^{R'} - \sum_k p_k S^R_k
\end{equation}
for {\em any} pure state ensemble for $\rho^{RQ'}$, and therefore
all of the elements of such an ensemble have $\rho^R_k = \rho^{R'}$.
Since we can consider any mixed state to be made up of pure states,
this is also true for $\rho^{RQ'}$ ensembles that include mixed states.

{\bf Fact 2:  Choice of ensemble is choice of ancilla measurement.}
Hughston, Jozsa and Wootters \cite{hjw} give a useful characterization of
all the pure state ensembles that can lead to a particular density operator
$\rho^A$ for a system $A$.  We ``purify'' the state by envisioning a pure
state $\ket{\psi^{AB}}$ of a larger composite system $AB$ such that
$\rho^A = \tr_B \proj{\psi^{AB}}$.  A measurement on system $B$ will
lead to an ensemble of relative states of $A$.  In \cite{hjw} it is
shown that, given a purification $\ket{\psi^{AB}}$ of $\rho^A$, we
can realize {\em any} ensemble for $\rho^A$ as an ensemble of relative
states for some measurement on $B$.  In other words, the choice of
$\rho^A$ ensemble is exactly the same as the choice of measurement
on the purifying system $B$.

In our context, we can include the environment system $E$ as before,
with the whole system $RQE$ in the pure state $\ket{\Psi^{RQE'}}$.
$E$ purifies $RQ$, so an ensemble of $RQ$ states corresponds to a
measurement on $E$.  From Fact 1, we know that every element of an
ensemble for $\rho^{RQ'}$ yields the same state $\rho^{R'}$ on $R$ alone.
Thus, for any possible outcome of any measurement
on $E$, the relative state of $R$ will be $\rho^{R'}$.
This means that the probabilities of the outcomes
of possible $R$-measurements are unaffected by the particular outcomes
of an $E$-measurement.

{\bf Fact 3:  No correlation implies product state.}
Quantum state tomography \cite{mikeandike}
allows the reconstruction of a quantum state
$\rho$ from the outcome distributions of a finite number of possible
measurements on the quantum system.  This procedure, when applied to a
composite quantum system $AB$, has two important features.  First, it is
sufficient to consider only product measurements of $A$ and $B$ to
do tomography of the joint state.  Second, if no statistical correlations
appear between the outcomes of the $A$ and $B$ measurements, the resulting
joint state must be a product state $\rho^A \otimes \sigma^B$.  Thus, a
necessary and sufficient condition for $A$ and $B$ to be in a product state
is that no correlations arise in any product measurement of the systems.

Since we have shown that $\entform = S^Q$ implies no statistical correlations
between $E$-measurements and $R$-measurements on the output state,
we can conclude that the output state
of the subsystem $RE$ is a product state $\rho^{R'} \otimes \sigma^{E'}$.
Given such a product state, we can apply the procedure in
\cite{sn96} to give an explicit error correction
operation $\errcorr$ that will restore the input state $\ket{\Psi^{RQ}}$
of $RQ$ with perfect fidelity.  Therefore, perfect quantum error correction
is possible if and only if $\entform = S^Q$.

\section{Intrinsic expressions for $I$ and $\entform$}

Both the coherent information $I$ and the entanglement of formation $\entform$
are ``intrinsic'' quantities to the system $Q$---that is, they can be expressed
entirely in terms of the input state $\rho^Q$ of $Q$ alone and the
trace-preserving CP map $\superop$ that describes $Q$'s dynamics.  First,
we note that the map $\superop$ can be given an ``operator sum'' representation
\cite{mikeandike}:
\begin{equation}
	\superop \left ( \rho^Q \right )  =
		\sum_{k} A_k \rho^Q A_k^{\dagger} ,
\end{equation}
where the $A_k$ operators satisfy $\bigsum{k} A_k^{\dagger} A_k = 1$. 
A given $\superop$ always has many different operator sum representations.
Suppose we have a unitary matrix $V_{kl}$, and define some operators $B_k$ as
linear combinations of the $A_k$'s:
\begin{equation}
	B_k = \sum_l V_{kl} \, A_l .
\end{equation}
Then the $B_k$'s give an alternate operator sum representation for $\superop$.

The operator sum representation is closely related to the
unitary representation for $\superop$, in which $\superop$ is given via
unitary evolution on a larger system that includes the environment $E$.
Once again, $E$ is taken to be initially in a pure state $\ket{0^E}$, and
the interaction of $Q$ and $E$ is given by the unitary operator $U^{QE}$.
Let $\ket{k^E}$ be a basis of $E$ states, and define the operator $A_k$
on $Q$ by the ``partial inner product''
\begin{equation}
	A_k \ket{\psi^Q} =  \bra{k^E} U^{QE} \ket{\psi^Q 0^E} ,
\end{equation}
where $\ket{\psi^Q 0^E}$ is shorthand for $\ket{\psi^Q} \otimes \ket{0^E}$.
We can use the $\ket{k^E}$ basis to do a partial trace over the $E$ system,
so that
\begin{eqnarray}
	\superop \left ( \rho^Q \right ) & = &
		\tr_E \, \left [ U^{QE} \left ( \rho^{Q} \otimes \proj{0^E} \right )
			{U^{QE}}^{\dagger} \right ] 	\nonumber \\
	& = & \sum_k \bra{k^E} U^{QE} \left ( \rho^{Q} \otimes \proj{0^E} \right )
			{U^{QE}}^{\dagger}  \ket{k^{E}}  \nonumber  \\
	& = & \sum_k  A_k \rho^Q A_k^{\dagger} .
\end{eqnarray}
The unitary freedom in the operator sum representation is the same as the
freedom to choose a basis for the environment system $E$.

The operator sum representation of $\superop$ gives the output state
$\rho^{Q'} = \superop(\rho^Q)$ as an ensemble of $Q$ states.  If we let
\begin{eqnarray}
	p_k & = & \tr A_k \rho^Q A_k^{\dagger}	\nonumber \\
	\rho^Q_k & = & \frac{1}{p_k} \left ( A_k \rho^Q A_k^{\dagger} \right ) ,
\end{eqnarray}
then $\rho^{Q'} = \bigsum{k} p_k \rho^Q_k$.  Different operator sum
representations yield different ensembles for the same output state.

The entanglement of formation $\entform$ of the $\rho^{RQ'}$ state
can be written
\begin{equation}
	\entform = \min \sum_k p_k S_k^Q
\end{equation}
where $S_k^Q$ is the entropy of $\rho_k^Q$ (as defined above) and
the minimum is taken over all operator sum representations for $\superop$.
In a similar way, the coherent information $I$ can be written
\begin{equation}
	I = S^{Q'} - \min H(\vec{p})
\end{equation}
where $H(\vec{p}) = - \bigsum{k} p_k \log p_k$ and the minimum is once again
taken over all operator sum representations \cite{s96}.
We can see why this is true
by appealing to a unitary representation.  The $p_k$'s are the diagonal
entries of the output density matrix for the environment $E$, and
$S^{E'} = \min H(\vec{p})$ (where we minimize over basis states).
Since the global state of $RQE$ is pure, $S^{E'} = S^{RQ'}$.

\section{Generalization}

We pointed out that both $I$ and $\entform$ were measures of entanglement
of the state $\rho^{RQ'}$, and that $I = \entform = S^Q$ for the input
pure state $\ket{\Psi^{RQ'}}$.  We now consider other possible
measures of the entanglement of $\rho^{RQ'}$.  Suppose $M$ is such a
measure, and that it satisfies the following conditions:
\begin{enumerate}
	\item $M = S^Q$ when $RQ$ is in a pure state.
	\item $M$ is additive if we have many copies of $\rho^{RQ'}$; that is,
		\begin{equation}
			M \left ( (\rho^{RQ'})^{\otimes n} \right ) =
				n \, M \left ( \rho^{RQ'} \right ) .
		\end{equation}
	\item  $M$ does not increase on average under local operations on,
		or classical communication between, $R$ and $Q$.
\end{enumerate}
Coherent information satisfies (1) and (2) but not (3); the asymptotic
entanglement of formation $\entform_{\infty}$ satisfies all three; it is
not known whether the ``single system'' entanglement of formation $\entform$
satisfies (2) (for this is exactly the question of whether $\entform =
\entform_{\infty}$).  Conditions (1)--(3) are similar to those discussed
in \cite{plenved}.

We will now show that perfect quantum error correction is possible
if and only if $M = S^Q$ for the output state $\rho^{RQ'}$.

``Only if'' is easy to see.  Initially, $M = S^Q$.  If $M$ decreases under the
action of $\superop$ on $Q$, then this loss cannot be made up by any error
correction procedure, which must be a local operation on $Q$.  Thus, the
original state can be restored only if $M = S^Q$ after $\superop$ acts.

To show that $M = S^Q$ is sufficient to allow perfect error correction, we
will show that $M \leq \entform$.  Imagine that we begin with $n \entform_{\infty}$
maximally entangled qubit pairs, for which $M_n = n \entform_{\infty}$.  We know
that, if $n$ is large, we can use these pairs to make about $n$ copies of our
state $\rho^{RQ'}$ by local operations and classical communication.  Since $M$
cannot increase in this process, $n M \leq M_n$, and so $M \leq \entform_{\infty}$.
But we have seen that $\entform_{\infty} \leq \entform$, so $M \leq \entform$.

We know that $\entform \leq S^Q$.  Thus, if $M = S^Q$ then $\entform = S^Q$.
As we have seen, this is sufficient to guarantee the existence of a perfect
error correction operation $\errcorr$ for $Q$.  $M = S^Q$ is therefore both
necessary and sufficient for the existence of $\errcorr$.

Remarkably, inequivalent entanglement measures lead to equivalent conditions
for perfect quantum error correction.
The coherent information $I$, the entanglement of formation $\entform$ (or its
asymptotic form $\entform_{\infty}$), and entanglement measures $M$ satisfying
our properties all share the feature that they are conserved by the evolution
$\superop$ on $Q$ only when that evolution produces no correlations between
$R$ and $E$.

\section{Remarks}

We have assumed that $Q$ may interact with environment, while $R$ remains
untouched.  Suppose instead that both $Q$ and $R$ independently interact
with separate parts of the environment, so that
\begin{equation}
	\rho^{RQ'} = \superop^R \otimes \superop^Q \left ( \proj{\Psi^{RQ}} \right ).
\end{equation}
We say in this case that perfect quantum error correction is possible if
the original state of $RQ$ can be restored by local operations and classical
communication.  It turns out that this can be done if and only if $\entform = S^Q$;
furthermore, if error correction is possible at all,
then no classical communication between $R$ and $Q$ is necessary.

Once again, $\entform = S^Q$ is plainly a necessary condition, and we must
show that it is also sufficient.  Suppose $\entform = S^Q$ after the operation
$\superop^R \otimes \superop^Q$.  We can imagine that this operation occurs in
two stages:
\begin{equation}
	\rho^{RQ'} = ( \superop^{R} \otimes I^Q ) \circ
			(  I^R \otimes \superop^Q ) \left (
			\proj{\Psi^{RQ}} \right ) .
\end{equation}
After the first stage, in which $I^R \otimes \superop^Q$ acts, we must have
$\entform = S^Q$.  Therefore, at this stage there exists an operation $\errcorr^Q$
on $Q$ that can accomplish perfect error correction.  That is,
\begin{equation}
	\proj{\Psi^{RQ}} = (I^{R} \otimes \errcorr^Q) \circ
			( I^R \otimes \superop^Q )
			\left ( \proj{\Psi^{RQ}} \right ) .
\end{equation}
Alternately, we note that
\begin{equation}
	\rho^{RQ'} = ( I^R \otimes \superop^Q ) \circ
			( \superop^{R} \otimes I^Q )
			\left ( \proj{\Psi^{RQ}} \right ) ,
\end{equation}
in which case $\entform = S^Q = S^R$ after the first operation, and an error
correction operation $\errcorr^R$ exists at this stage:
\begin{equation}
	\proj{\Psi^{RQ}} = ( \errcorr^{R} \otimes I^Q ) \circ
			( \superop^R \otimes I^Q )
			\left (	\proj{\Psi^{RQ}} \right ) .
\end{equation}
Now we can see that $\errcorr^R \otimes \errcorr^Q$ will correct the complete
operation:
\begin{eqnarray}
	\lefteqn{(\errcorr^R \otimes \errcorr^Q) \circ ( \superop^R \otimes \superop^Q )
		\left ( \proj{\Psi^{RQ}} \right ) }   \nonumber \\
	&  =  &
	( I^R \otimes \errcorr^Q) \circ ( \errcorr^R \otimes I^Q )
	\circ ( I^R \otimes \superop^Q ) \circ ( \superop^R \otimes I^Q )
		\left ( \proj{\Psi^{RQ}} \right ) 	\nonumber \\
	& = &
	( I^R \otimes \errcorr^Q) \circ ( I^R \otimes \superop^Q )
	\circ ( \errcorr^R \otimes I^Q ) \circ ( \superop^R \otimes I^Q )
		\left ( \proj{\Psi^{RQ}} \right ) 	\nonumber \\
	& = & \proj{\Psi^{RQ}} .
\end{eqnarray}
Thus, $\entform = S^Q$ is a necessary and sufficient condition for local
correction of the quantum state, even if both subsystems have
experiened independent noisy evolutions.

Throughout this paper, we have focused our attention on the issue of
{\em perfect} error correction.  What about {\em approximate} error
correction?  We have elsewhere \cite{approxec} shown that, if the
loss of coherent information is small, then an operation $\errcorr$
exists that will nearly restore the original state $\ket{\Psi^{RQ}}$.
To be precise, if $S^Q - I < \epsilon$, then there exists an operation
$\errcorr$ on $Q$ that will restore the input state with fidelity
$F > 1 - 2 \sqrt{\epsilon}$.  Is there an analogous theorem for the
entanglement of formation $\entform$?  That is, suppose $S^Q - \entform
< \epsilon$.  With what fidelity can error correction be performed?
This and many other questions remain unresolved.

We are happy to acknowledge very useful discussions with C. H. Bennett
and J. A. Smolin that clarified these results.

\end{document}